\documentclass{article}
\usepackage{graphicx} 
\usepackage{amssymb}
\usepackage{amsmath}
\usepackage{amsthm}
\usepackage{xcolor}
\usepackage{ulem}
\usepackage{comment}
\usepackage{cite}

\usepackage{fullpage}
\usepackage[colorlinks=true]{hyperref}

\newtheorem{theorem}{Theorem}[section]

\theoremstyle{definition}
\newtheorem{remark}[theorem]{Remark}

\newcommand{\bra}[1]{\langle#1| }
\newcommand{\ket}[1]{|#1\rangle }

\newcommand{\h}{\mathfrak{h}}
\newcommand{\Tr}{\text{Tr}}

\newcommand{\F}{\mathcal{F}}
\newcommand{\U}{\mathcal{U}}

\newcommand{\atgh}{\mathrm{artanh}}

\title{Exponential concentration of fluctuations in mean-field boson dynamics}
\author{Matias Gabriel Ginzburg \thanks{Scuola Internazionale Superiore di Studi Avanzati, Via Bonomea, 265, 34136 Trieste TS, Italy}, Simone Rademacher \thanks{Department of Mathematics, LMU Munich, Theresienstr. 39, 80333 Munich, Germany} and Giacomo De Palma \thanks{University of Bologna, Department of Mathematics, Piazza di Porta San Donato 5, 40126 Bologna BO, Italy}}

\begin{document}

\maketitle

\begin{abstract}
We study the mean-field dynamics of a system of $N$ interacting bosons starting from an initially condensated state. For a broad class of mean-field Hamiltonians, including models with arbitrary bounded interactions and models with unbounded interaction potentials, we prove that the probability of having $n$ particles outside the condensate decays exponentially in $n$ for any finite evolution time.
Our results strengthen previously known bounds that provide only polynomial control on the probability of having $n$ excitations.
\end{abstract}

\section{Introduction}

In this paper, we study the dynamics of a large number $N$ of bosons in the mean-field regime, which arises in various areas of physics, including spin systems \cite{H,S}, and Bose-Einstein condensates \cite{S,RS}. In typical experimental set-ups, for instance in experiments on Bose-Einstein condensation \cite{Wieman,Ketterle}, the initial state is prepared in a condensed phase, meaning that a macroscopic fraction of the $N$ particles occupies the same one-particle quantum state, referred to as the condensate. It is well established, both mathematically and physically, that the mean-field dynamics preserves condensation \cite{H,S,RS}.

We contribute to the analysis of condensation along the mean-field dynamics by proving that the probability of having $n$ particles outside the condensate decays exponentially with $n$ for any finite evolution time. This result strengthens previously known bounds, which typically prove only a polynomial decay of such a probability.

To be more precise, let $\mathrm{Sym}^N \mathfrak{h}$ be the $N$-fold symmetric tensor product of the single-particle Hilbert space $\mathfrak{h}$. We describe the dynamics of the $N$ bosons by a wavefunction $\Psi_N(t)\in \mathrm{Sym}^N \mathfrak{h}$ solving the Schr\"odinger equation

\begin{equation}\label{eq:SchrodingerEq}
  i\partial_t \Psi_N(t)= H_N \Psi_N(t),
\end{equation}
where the mean-field Hamiltonian $H_N$ is given by
\begin{equation}\label{def:Mean_field_H}
  H_N = \sum_{i=1}^N T_i + \frac{1}{N-1}\sum_{1\le i<j\le N} w_{ij}.
\end{equation}
Here, $T_i$ denotes the self-adjoint $N$-particle operator acting as the self-adjoint single-particle operator $T:\mathfrak{h}\to\mathfrak{h}$ on the $i$-th particle and as the identity on all the other particles. Similarly, $w_{ij}$ denotes the self-adjoint two-body interaction acting as the self-adjoint operator $w:\mathfrak{h}^{\otimes 2}\to \mathfrak{h}^{\otimes 2}$ on the particles $i$ and $j$ and as the identity on all the remaining particles\footnote{If $T$ or $w$ are unbounded, their domain is a dense subspace of $\mathfrak{h}$ or $\mathfrak{h}^{\otimes2}$ rather than the full $\mathfrak{h}$ or $\mathfrak{h}^{\otimes2}$, respectively}. In accordance with bosonic symmetry, we assume that $w$ is invariant under permutation of the two particles.

We consider two physically relevant classes of mean-field models:
\begin{itemize}
\item[(i)] \emph{Arbitrary bounded interactions:} we assume that the two-body interaction $w$ is bounded, i.e. $\|w\|<\infty$, but otherwise arbitrary.
\item[(ii)] \emph{Unbounded interaction potentials:} we set $\mathfrak{h}=L^2(\mathbb{R}^3)$, $T=-\Delta$, and assume that the interaction is given by multiplication with a potential of the form $w(x,y)=V(x-y)$, where $V$ satisfies the operator inequality $V^2\le D(1-\Delta)$ for some constant $D>0$.
\end{itemize}

Mean-field models of type (i) naturally arise, for example, in the description of spin systems \cite{H,S}, and Lipkin-Meshkov-Glick-type models \cite{LMG}. In the continuum, bounded interactions arise naturally as soft-core or regularized potentials of finite height, which are used to model effective interactions in ultracold atomic gases, for instance in Rydberg-dressed systems and related proposals for supersolid phases \cite{Cinti2010DropletCrystalSupersolid}. The Kac model and soft spheres can also be described by type (i) models \cite{paul2019size}.

Mean-field models with unbounded interaction potentials of type (ii) constitute a standard mathematical framework for the description of Bose--Einstein condensation, and the condition $V^2\leq D(1-\Delta)$ is chosen in order to include the Coulomb potential; see for example \cite{RS}. In typical experimental realizations \cite{Wieman,Ketterle}, dilute gases of bosonic atoms are confined in external traps and cooled to ultra-low temperatures. Below a critical temperature, a macroscopic fraction of the particles occupies the same one-particle quantum state, giving rise to Bose-Einstein condensation.

Motivated in this experimental realization, we study the many body dynamics \eqref{eq:SchrodingerEq} for initial data given by a condensate state
\begin{equation}\label{eq:initialCondensate}
    \Psi_N(0)=\sum_{n=0}^{N} \phi(0)^{\otimes(N-n)} \otimes_s \xi_n
\end{equation}
with $\phi(0)\in\h$ the initial condensate wavefunction, and $\xi_n$ are $n$-particle wavefunctions describing the excitations. If the initial number of excitations is exponentially controlled, in a sense that we will make formal later, we will show that the property of  exponential condensation is preserved for positive times. More precisely, the condensate wavefunction $\phi(t)$ , satisfying $\| \phi (t) \|_2 =1$, is expected to evolve according to the Hartree equation

\begin{equation}\label{eq:Hartree}
  i\partial_t \phi(t)= H_H^{\phi(t)}\phi(t), \qquad H_H^{\phi (t)} = T + w^{\phi (t)},
\end{equation}
where the effective one-particle operator $w^{\phi (t)}$ is defined through
\begin{equation}
  \langle \varphi_1, w^{\phi (t)} \varphi_2\rangle
  = \langle \varphi_1\otimes \phi (t),\, w\, \varphi_2\otimes \phi (t) \rangle,
  \qquad \forall\, \varphi_1,\varphi_2\in\mathfrak{h}.
\end{equation}
If $w$ is bounded, the Hartree equation \eqref{eq:Hartree} always has a unique mild solution defined for any $t\in\mathbb{R}$ from \cite[Chapter 6, Theorem 1.4]{Pazy1992}.

Particles outside the condensate are counted by the excitation number operator
\begin{equation*}
\label{def:N+}
  \mathcal{N}_+(t)=\sum_{i=1}^N Q(t)_i,
\end{equation*}
where $Q(t)_i$ denotes the operator acting as the orthogonal projector
$Q(t)=1-|\phi(t)\rangle\langle \phi(t)|$ on the $i$-th particle and as the identity on all others.

Mathematically, condensation at positive times is expressed by
\begin{equation}
  \frac{1}{N}\langle \Psi_N(t), \mathcal{N}_+(t)\Psi_N(t)\rangle \to 0
  \qquad \text{as } N\to\infty.
\end{equation}
This property is well understood for models of type (ii) under suitable assumptions on the interaction potential and the initial data \cite{GV3,GV4,GV2,H,S,BGM,EY,RS,KP,P,ES,CLS,CL,L,CLS,DL}. More recently, these results have been refined by proving uniform bounds on higher moments of the excitation number,
\begin{equation}
  \langle \Psi_N(t), (\mathcal{N}_+(t))^k \Psi_N(t)\rangle = \mathcal{O}(1),
\end{equation}
for fixed $k\in\mathbb{N}$ in the limit $N\to\infty$; see, for example,
\cite{BPPS,LNS,NM1,NM2,MPP,GM1,GM2,GMM1,GMM2}.

These bounds imply that the probability of finding more than $n$ particles outside the condensate decays at least polynomially in $n$. In this work, we strengthen these results by proving exponential decay of this probability; see \autoref{rmk:prob} below for a precise formulation.

\section{Exponential condensation}

 In this section we present our results on exponential condensation, more precisely we consider an initial condensate state \eqref{eq:initialCondensate} satisfying
\begin{equation}\label{eq:initial_condensate_condition}
    \langle \Psi_N(0), \exp(\beta\mathcal{N}_+(0))\Psi_N(0)\rangle \leq C_\beta
\end{equation}
for some $C_\beta>0$ independent of $N$. Then in the large $N$ limit

\begin{align}
\label{eq:exp-N}
\langle \Psi_N (t), \exp \big( \beta \mathcal{N}_+(t) \big) \Psi_N (t) \rangle  = \mathcal{O}(1) 
\end{align}
for $\beta \leq \beta_c (t)$ and suitable $\beta_c (t) >0$ discussed below for the two different types of models (i), (ii). The result for mean-field models of type (i) is given in \autoref{theo:ExpCondansation} in \autoref{sec:theo_bounded}, and the result for mean-field models of type (ii) in \autoref{theo:ExpCondansation_coulomb} in \autoref{sec:theo_coulomb}.

\subsection{Mean-field models with bounded potentials}\label{sec:theo_bounded}

We recall that in this section, we study mean-field models of type (i) arising for example  in the context of quantum spin systems.

\begin{theorem}[Exponential Condensation for bounded potentials]\label{theo:ExpCondansation} Let $w$ be an arbitrary bounded two-body interaction, let $\Psi_N(t)\in\mathrm{Sym}^N\h$ denote the solution to the Schr\"odinger equation \eqref{eq:SchrodingerEq} with initial data  \eqref{eq:initialCondensate} satisfying the condensate condition \eqref{eq:initial_condensate_condition}. Furthermore, let $\phi (t)$ denote the solution to the Hartree equation \eqref{eq:Hartree} with initial data $\phi (0) \in \h$.   

Then, the number of excitations $\mathcal{N}_+(t)$ defined by \eqref{eq:Num_+}, satisfies 
\begin{equation}
\label{eq:bound-exp-N}
  \left\langle \Psi_N(t), \exp(\beta \mathcal{N}_+(t))\;\Psi_N(t)\right\rangle\leq  C_\beta f(t,\beta)
\end{equation}
for all $t\geq 0$ and $0\leq \beta < \beta_c(t) = -\ln\tanh(3\|w\|\,t)$, where
\begin{equation}
    f(t,\beta) =  \left( \frac{1-  \tanh\left(3\|w\|\,t\right) e^{-\beta}}{1- \tanh\left(3\|w\|\,t\right) e^{\beta} }\right)^{1/3}\,.
\end{equation}
\end{theorem}

\begin{remark}\label{rmk:f,beta}
We collect a few remarks on the constants $f(t,\beta)$ and $\beta_c(t)$ appearing in \autoref{theo:ExpCondansation}.
\begin{itemize}
\item[(i)] Since the bound $f(t,\beta)$ on the moment generating function of $\mathcal{N}_+(t)$ does not depend on the total number of particles $N$, \autoref{theo:ExpCondansation} indeed establishes exponential condensation in the sense of~\eqref{eq:exp-N}. Moreover, $f(0,\beta)=1$ for all $\beta>0$ and $f(t,0)=1$ for all $t\geq0$. Hence, for the pairs $(0,\beta)$ and $(t,0)$, the upper bound provided by the theorem coincides with the exact value of the corresponding quantity.
\item[(ii)] We note that $\beta_c(t)>0$ for all finite times, and therefore exponential condensation holds for any $t\geq0$. The critical parameter $\beta_c(t)$ scales as $\mathcal{O}(\log(t^{-1}))$ for small times and as $\mathcal{O}(e^{-t})$ for large times. In particular, $\beta_c(t)$ becomes exponentially small as $t\to\infty$.
\end{itemize}
\end{remark}

\begin{remark}[Probabilisitc picture]  \label{rmk:prob} As mentioned above, the number of excitations in the state $\Psi_N(t)$ can be interpreted as a random variable with distribution
\begin{align}
\mathbb{P}_{\Psi_N(t)}\big[\mathcal{N}_+(t) = n\big]
= \langle \Psi_N(t), \mathbf{1}_{\{\mathcal{N}_+(t)=n\}}\, \Psi_N(t) \rangle .
\end{align}
In this probabilistic picture, \autoref{theo:ExpCondansation} provides an upper bound on the moment generating function of $\mathcal{N}_+(t)$. As a consequence of Markov's inequality, we obtain
\begin{align}
\mathbb{P}_{\Psi_N(t)}\big[\mathcal{N}_+(t) > n\big]\leq 
C_\beta f(t,\beta)\, e^{-\beta n}
\end{align}
for all $\beta<\beta_c(t)$. Hence, the probability of finding more than $n$ particles outside the condensate decays exponentially in $n$.

\end{remark}

\subsection{Mean-field models with unbounded potentials }\label{sec:theo_coulomb}

We recall that in this section, we study mean-field models of type (ii) used as a mathematical framework for the analysis of Bose-Einstein condensates. Thus, in this section we study the mean-field Hamiltonian 
\begin{equation}\label{eq:mean_field_H_multiplication}
    H_N = \sum_{i=1}^{N}-\Delta_{x_i} + \frac{1}{N-1} \sum_{1\leq i < j \leq N}V(x_i-x_j) 
\end{equation}
and assume that the interaction potential $V:\mathbb{R}^3\to\mathbb{R}$ satisfies the operator inequality 
\begin{equation}\label{eq:V^2_bound}
    V(x)^2 \leq D(1-\Delta_x)
\end{equation}
for some $D>0$, i.e. 
\begin{equation}
\int_{x\in\mathbb{R}^3}V(x)^2\left|\psi(x)\right|^2 dx \le D \int_{x\in\mathbb{R}^3}\left(\left|\psi(x)\right|^2 + \left|\nabla\psi(x)\right|^2\right)dx\,.
\end{equation}
for any $\psi\in H^1(\mathbb{R}^3)$. The associated Hartree dynamics \eqref{eq:Hartree} has a unique mild solution, and, moreover 
\begin{equation}\label{def:K}
    \mathbb{V}:= \sup_x \int |V(x-y)\phi (t,y)|^2dy < \infty , 
\end{equation}
see for example \cite{RS}. In this setting, we prove exponential condensation of the form \eqref{eq:exp-N}, too. 

\begin{theorem}[Exponential condensation for unbounded interaction potentials]\label{theo:ExpCondansation_coulomb}
Let $\h = L^2   (\mathbb{R}^3)$, let $V$ satisfy \eqref{eq:V^2_bound}, let $\Psi_N(t)\in\mathrm{Sym}^N\h$ denote the solution to the Schr\"odinger equation\eqref{eq:SchrodingerEq} with initial datum \eqref{eq:initialCondensate} satisfying the condensate condition \eqref{eq:initial_condensate_condition}. Furthermore, let $\phi (t)$ denote the solution to the Hartree equation \eqref{eq:Hartree} with initial datum $\phi (0) \in \h$.   

Then, the number of excitations $\mathcal{N}_+$ defined by \eqref{eq:Num_+} satisfies 
\begin{equation}
  \left\langle \Psi_N(t), \exp(\beta \mathcal{N}_+(t))\;\Psi_N(t)\right\rangle\leq C_\beta f(t,\beta)
\end{equation}
for  all $t \geq 0$ and for all $0\leq \beta < \beta_c(t) = -\ln\tanh(6\mathbb{V}\,t)$, where
\begin{equation}
    f(t,\beta) = \left( \frac{1-  \tanh\left(6\mathbb{V}\,t\right) e^{-\beta}}{1- \tanh\left(6\mathbb{V}\,t\right) e^{\beta} }\right)^{1/3}\,.
\end{equation}
\end{theorem}

\subsubsection{Comparison to the literature}
As discussed in the introduction, our results fit into a broad body of literature on the analysis of Bose--Einstein condensation for mean-field models \cite{GV3,GV4,GV2,H,S,BGM,EY,RS,KP,P,ES,CLS,CL,L,CLS,DL}, as well as on the study of quantum fluctuations, i.e.\ particles outside the condensate, in such systems \cite{BPPS,LNS,NM1,NM2,MPP,GM1,GM2,GMM1,GMM2}; see also the review article \cite{N}. Our results, however, complement the existing literature on mean-field dynamics by providing a stronger, exponential control of the number of excited particles. We remark that, for large times, the critical value $\beta_c$ decays exponentially in time, and our bounds are consistent with earlier results on moment estimates for the number of excitations, which typically grow exponentially in time; see, for example, \cite{RS} under similar assumptions on the initial data and interaction potential.

\subsubsection{Probabilistic picture} To the best of our knowledge, our work establishes the first bound on the moment generating function of the number of excitations along the many-body time evolution; see \autoref{rmk:prob}. For mean-field models in the ground state, bounds on the moment generating function have been obtained earlier \cite{MP,NR,BBR}, and its asymptotic behavior has been analyzed more recently in \cite{R2}. From this perspective, our results provides a first step towards exponential control of excitations in the time-dependent setting.

\subsubsection{Singular scaling regimes}
From a physical point of view, the most relevant but mathematically most challenging setting is the analysis of singular interaction potentials in the so-called Gross-Pitaevskii regime. In this regime, the mean-field interaction potential $V$ is replaced by an $N$-dependent potential $V_N = N^3 V(N\cdot)$, which scales with the total number of particles and converges, in the large-particle limit, to a $\delta$-interaction. Remarkably, bounds \cite{NR,BBR,R2} and asymptotic formulas for the moment generating function of the number of excitations in the ground state have been shown to remain valid even in this mathematically challenging regime. For the corresponding many-body dynamics, however, only bounds on the first moment of the number of excitations are currently known \cite{BS}, while bounds on higher moments remain a challenging open problem.

\subsection{Idea of the proofs}

We prove both theorems in \autoref{sec:Proof} based on the same idea:  The goal is to derive a bound for the function $g_N : \mathbb{R}_+ \times \mathbb{R}_+ \rightarrow \mathbb{R}$ by 
\begin{equation}\label{def:g_}
    g_N(t,\beta) =\left\langle \Psi_N(t), \exp(\beta \mathcal{N}_+(t))\Psi_N(t)\right\rangle 
\end{equation}
by a Gronwall-type argument. We will first compute the time derivative of $\partial_tg_N$ and show in a second step that we can bound it from above in terms of $g_N$ and $\partial_\beta g_N $. In the last step we then perform a Gronwall type argument and arrive at the desired bound. 

Before presenting the proof in \autoref{sec:Proof}, in the following Section we first give a brief introduction to the Fock space formalism and the so-called excitation map, introduced by \cite{LNSS}, that our analysis relies on. 

\section{Excitation map}
\label{sec:second_quantization}

In this section we introduce the excitation map, introduced in \cite{LNSS}, defined for any $N$-particle wavefunction to factor out any contributions of the Hartree evolution $\phi (t)$. 

The excitation map is based on the idea that while the one-particle Hilbert space $\h$ has a decomposition into $\h = \text{span}(\{\phi(t)\}) \oplus \text{span}(\{\phi(t)\})^\perp$, the $N$-particle Hilbert space $\mathrm{Sym}^N\h$ has a decomposition into products of $\text{span}(\{\phi(t)\})$ and 
\begin{equation}
    \F_{ \perp \phi(t)}^{\leq N} := \bigoplus_{n=0}^{N}\mathrm{Sym}^n\h_+(t) \; , 
\end{equation}
with $\h_+(t) = \text{span}(\{\phi(t)\})^\perp$ called the orthogonal or excitation Fock space. In contrast to the standard bosonic Fock space 
\begin{equation}
    \F := \bigoplus_{n=0}^{\infty}\mathrm{Sym}^n\h, 
\end{equation}
the orthogonal Fock space is defined over $\h_+(t)$ (instead of $\h$) and contains wavefunctions of at most $N$ particles. On the Fock space $\F$, we have the usual creation and annihilation operators given by 
\begin{align}
a_i^* = a (e_i), \quad \text{and} \quad a_i = a(e_i) 
\end{align}
where we set $e_0 := \phi(t)$ and $\{e_i\}_{i\in \mathbb{N}_+}$ as an orthonormal basis of $\text{span}(\{\phi(t)\})^\perp$. We remark that to simplify notation, we neglect the time dependence of the basis $\{e_i\}_{i\in \mathbb{N}_+}$ and $e_0$ in their notation. The creation and annihilation operators satisfy the standard canonical commutation relations 
\begin{align}
[a_i^*,a_j^*] = [a_i,a_j] =0, \quad \text{and} \quad [a_i,a_j^*]  =\delta_{ij}\; . 
\end{align}

For the definition of the excitation map, we embed any $N$-particle wavefunction into the bosonic Fock space through  $W_N :\mathrm{Sym}^N\h \rightarrow \mathcal{F} $ given by 
\begin{equation}\label{def:W_N}
    W_N(\Psi_N) :=  \left(\bigoplus_{n=0}^{N-1}0\right) \oplus\Psi_N \oplus \left(\bigoplus_{n=0}^{N-1} 0\right), \quad \text{and} \quad  W_N^*\left(\bigoplus_{n=0}^\infty \psi_n \right) = \psi_N 
\end{equation}
where $W_N^*$ denotes the adjoint of $W_N$. In particular, note that while $W_N^*W_N$ acts as the identity on $\mathrm{Sym}^N\h$, the adjoint $W_NW_N^*$ acts on $\F$ as a projector onto the subspace of $N$ particles.  Moreover, we define $Q_{N,t}:\F\to\F$ as the orthogonal projector from the full bosonic Fock space $\mathcal{F}$ on the orthogonal Fock space $\mathcal{F}_{\perp \phi (t)}^{\leq N}$ by

\begin{equation}\label{def:Q_N}
    Q_{N,t} := \left(\bigoplus_{n=0}^{N} (1-\ket{\phi(t)}\bra{\phi (t)})^{\otimes n} \right) \oplus \left(\bigoplus_{n=N+1}^{\infty} 0 \right) .
\end{equation}
Then, the excitation map $U_{N,t}: \mathrm{Sym}^N\h \to \F$ \cite{LNSS}, is given by 
\begin{equation} \label{eq:excitationMap}
    U_{N,t} = \sum_{n=0}^{N} Q_{N,t} \frac{a(\phi(t))^{N-n}}{\sqrt{(N-n)!}} W_N .
\end{equation}

Note that in this picture, the number of excitations defined in \eqref{def:N+} reads 
\begin{align}
\label{eq:Num_+}
  W_N  \mathcal{N}_+ (t) W_N^*  := \mathcal{N} - a^*(\phi(t))a(\phi(t)) = \sum_{i \in \mathbb{N}_+} a_i^*a_i \,,
\end{align}
where $\mathcal{N} = \sum_{i \in \mathbb{N}} a_i^*a_i$. 

By construction, the excitation map $U_{N,t}$ maps any $N$-particle wavefunction into the orthogonal Fock space by destroying all particles in the Hartree wavefunction $e_0 = \phi( t)$. In particular note that the image of $U_{N,t}$ is the orthogonal Fock space $\Im(U_{N,t})=\Im(Q_{N,t}) = \F^{\leq N}_{\perp \phi (t)}$, respecting the heuristics that an $N$-particle wavefunction can have at most $N$ excitations outside the Hartree evolution.  We can then define the initial excitations state
\begin{equation}\label{def:initialExcitations}
    \Xi_N := U_{N,0}\Psi_N(0) = \bigoplus_{n=0}^{N} \xi_n. 
\end{equation}

Furthermore, the conjugate of the excitation map is given by 
\begin{equation}
    U_{N,t}^* = \sum_{n=0}^{N} W_N^* \frac{(a ( \phi (t))^*)^{N-n}}{\sqrt{(N-n)!}} Q_{N,t} .
\end{equation}
Since $ U_{N,t}^*U_{N,t} = 1$  and $U_{N,t}U_{N,t}^* = Q_{N,t}$, the excitation map is a partial isometry.

The excitation map comes with nice properties (see also \cite{LNSS}): It transforms the projected creation operator to 
  \begin{equation}\label{eq:U_NConjugation}
    U_{N,t}W_N^* a_n^*W_{N-1}U_{N-1,t}^* = \begin{cases}
    Q_{N,t}\ a_n^*\ Q_{N-1,t}  & \forall\; n\geq 1\\
    Q_{N,t}\ \sqrt{N-\mathcal{N}}\ Q_{N-1,t} &n=0
    \end{cases} \; . 
\end{equation} 
Thus, in the limit of infinitely many particles, where we expect that $\mathcal{N} = O(1)$, a particle created through $a_ 0 = a ( \phi (t))$ along the Hartree evolution is effectively replaced by an $O(\sqrt{N})$ quantity, while excitations orthogonal to the Hartree evolution, created through $a_n^*$ with $n\geq 1$, are replaced by $O (1)$ quantities. In physics this formal transformation is referred to as c-number substitution. Note that $U_{N,t}$ furthermore maps the projected number of excitations \eqref{eq:Num_+} to the number of particles operator on the orthogonal Fock space, i.e. 

\begin{equation}\label{eq:Num_+ToNum}
\begin{split}
    U_{N,t} \mathcal{N}_+(t) U_{N,t}^* =& \sum_{n\geq1}U_{N,t} W_N^* a_n^*W_{N-1}U_{N-1,t}^*U_{N-1,t} W_{N-1}^* a_n W_{N}U_{N,t}^* = \mathcal{N} Q_N \;, 
\end{split}
\end{equation} 
where we used that $W_{N-1}U_{N-1,t}^*U_{N-1,t} W_{N-1}^* = 1$ on the image of $a_nW_NU_{N,t}^*$, similarly $Q_{N-1,t} = 1$ on the image of $a_n Q_{N,t}$ and $Q_{N,t}$ commutes with $\mathcal{N}$. The projected particle-number operator $\mathcal{N} = \sum_{i=0}^\infty a_i^*a_i$ (that is actually time independent) is mapped by the excitation map to 
\begin{equation}\label{eq:NumToN}
    U_{N,t} W_N^* \mathcal{N}\ W_NU_{N,t}^* = \mathcal{N} Q_{N,t}+(N-\mathcal{N}) Q_{N,t} = N Q_{N,t}. 
\end{equation}

For the proof of the theorems, we embed the $N$-particle wavefunction through $W_N$ in the bosonic Fock space. For the analysis, we therefore also need to transform the Hamiltonian $H_N$ orignially defined in \eqref{def:Mean_field_H} on the $N$-particle Hilbert space $\mathrm{Sym}^N\h$ to an operator on the Fock space: We recall that $H_N$ is defined as a sum of one-particle operators $T_i$ and two-particle operators $w_{ij}$ whose second quantization is defined as 
\begin{equation}
    d\Gamma(T) = \bigoplus_{n=1}^{\infty}\sum_{i=1}^{n} T_i, \quad d\Gamma(w) = \bigoplus_{n=2}^{\infty}\sum_{i<j} w_{ij} \; . 
\end{equation}
Thus, the $N$-particle Hamiltonian and the Hartree Hamiltonian act on the Fock space as 
\begin{equation}\label{eq:dGamma(H_N)}
    d\Gamma(H_{N}) = d\Gamma(T) + \frac{1}{N-1} d\Gamma(w),\quad  \text{resp.} \quad  d\Gamma(H_{H}^{\phi (t)} ) = d\Gamma(T) + d\Gamma(w^{\phi (t)}).
\end{equation} 
We remark that the interaction potential can be written in terms of creation and annihilation operators
\begin{equation}
    d\Gamma(w) = \frac{1}{2}\sum_{m,n,p,q\in\mathbb{N}} w_{mnpq} a^*_m a^*_n a_q a_p, \quad \text{and} \quad  d\Gamma(w^{\phi (t)})= \sum_{m,p\in\mathbb{N}}  w_{m0p0}  a^*_m\,,
\end{equation}
where $w_{mnpq} =\left\langle e_m\otimes e_n, w\;e_p\otimes e_q\right\rangle$, recalling that, in abuse of notation we set $\phi (t) = e_0$, and the series converges strongly in any subspace of the Fock space with bounded number of particles.  Furthermore, the matrix elements $w_{mnpq}$ satisfy the following properties: 
\begin{enumerate}
    \item[(i)] If $w$ is Hermitian, then $ w_{mnpq}^* = w_{pqmn}$. 
    \item[(ii)] If $w$ is symmetric under the exchange of particles, then $w_{mnpq} = w_{mnqp} = w_{nmpq}=w_{nmqp}$. 
\end{enumerate}

\section{Proof of the theorems}\label{sec:Proof}

In this section, we prove \autoref{theo:ExpCondansation} and \autoref{theo:ExpCondansation_coulomb}. The idea for both proofs is the same: The goal is to prove that the function $g_N : \mathbb{R}^2 \rightarrow \mathbb{R}$ given by 
\begin{equation}\label{def:g}
    g_N(t,\beta) =\left\langle \Psi_N(t), \exp(\beta \mathcal{N}_+ (t))\ \Psi_N(t)\right\rangle.
\end{equation}

satisfies a Gronwall type estimate of the form 
\begin{equation}\label{eq:general_bound_dtg}
    \partial_t g_N(t,\beta) \leq  \left( 8K\sinh\left(\frac{\beta}{2}\right) + 2K \sinh(\beta)\right) \partial_\beta g_N(t,\beta) + 2\sinh(\beta) K g_N(t,\beta).
\end{equation} 
where the constant $K>0$ depends on the choice of the two-particle operator $w$ in \autoref{theo:ExpCondansation_coulomb} resp. \autoref{theo:ExpCondansation}. In fact, we shall prove that for the choice  of bounded interactions, in \autoref{theo:ExpCondansation}, we have $K=\|w\|$, while for  unbounded interations, in \autoref{theo:ExpCondansation_coulomb}, we have $K= 2\mathbb{V}$.

For this we split the proof in three steps: In the first step in \autoref{sec:step1}, we compute the derivative of $\partial_t g_N (t,\beta)$ that works the same for both Theorems. In the second step, in \autoref{sec:step2}, we then derive an estimate of the form \eqref{eq:general_bound_dtg}. As discussed before, the final bound depends on the choice of the two-particle interaction $w$ and is therefore proven differently for both Theorems in \autoref{sec:bound_bounded_potentials} and \autoref{sec:bounds_coulomb_potential}. In the final third step in \autoref{sec:step3} we then perform a Gronwall type argument to derive the final results in \autoref{theo:ExpCondansation} and \autoref{theo:ExpCondansation_coulomb} from \eqref{eq:general_bound_dtg}.

\subsection{Step 1}\label{sec:step1}

In this section, we derive an explicit expression for the time derivative of $g_N (t,\beta)$ defined in \eqref{def:g}. Note that in fact, in the definition of $g_N (t,\beta)$ there are two quantities that depend on time: the many-body wavefunction $\Psi_N(t)$ and (in abuse of notation) the orthogonal number operator $\mathcal{N}_+(t)$. However, introducing the fluctuation dynamics 
\begin{align}
    \mathcal{U}_N(t,0) := U_{N,t} e^{-iH_N t}U_{N,0}^* 
\end{align}
and using that by our choice of initial data \eqref{def:initialExcitations}, we have from \eqref{eq:SchrodingerEq} and the identity \eqref{eq:Num_+ToNum} 

\begin{equation}\label{eq:g_U(t,0)}
    g_N(t,\beta) =\left\langle \mathcal{U}_N(t,0)\Xi_N, \exp(\beta\,\mathcal{N})\mathcal{U}_N(t,0)\Xi_N \right\rangle\; .  
\end{equation}
Thus, we passed the time dependency of the expectation value to the fluctuations dynamics $\mathcal{U}_N (t;0)$ only. The fluctuation dynamics satisfies 
\begin{equation}
    i \partial_t \U_N(t,0) = \mathcal{L}_N(t) \U_N (t,0) ,
\end{equation}
with the evolution generator 
\begin{equation} \label{eq: L(t)}
    \mathcal{L}_N(t) = (i\partial_t U_{N,t})U_{N,t}^* + U_{N,t} H_N U_{N,t}^*\,.
\end{equation}
In the following, we use the short-hand notation $\left\langle \cdot\right\rangle$ for the expectation value of any operator in the state $\U_N(t, 0) \Xi_N$. With this notation, the time derivative of $g_N$ reads 
\begin{equation}\label{eq:parital_tg_N}
    \partial_t g_N(t,\beta) = -i \left\langle [\exp(\beta\mathcal{N}),\mathcal{L}_N(t)] \right\rangle. 
\end{equation}
To calculate the commutator, we first derive an explicit expression for the generator $\mathcal{L}_N (t)$ of the fluctuation dynamics \eqref{eq: L(t)} in terms of creation and annihilation operators. We start with the time derivative of the excitation map. We recall the definition of $U_{N,t}$ in \eqref{eq:excitationMap}. Since 
\begin{equation}\label{eq:partial_t a_0}
   i\partial_t a ( \phi(t)) = [d\Gamma(H_{H}^{\phi (t)}),a ( \phi (t))], \quad \text{and} \quad i\partial_t Q_{N,t} = [d\Gamma(H_{H}^{\phi (t)}),Q_{N,t}],
\end{equation}
we find 
\begin{equation}
\label{eq:U_ndt}
\begin{split}
    i\partial_t U_{N,t} &= \left[d\Gamma(H_{H}^{\phi(t)}),\sum_{n \in \mathbb{N}} Q_{N,t} \frac{a( \phi(t))^{N_n}}{\sqrt{(N-n)!}} \right] W_N \\
    &= d\Gamma(H_{H}^{\phi (t)}) U_{N,t} - \left(\sum_{n \in \mathbb{N}} Q_{N,t} \frac{a ( \phi (t))^{N_n}}{\sqrt{(N-n)!}}\right) d\Gamma(H_{H}^{\phi (t)}) W_N \\
    &= d\Gamma(H_{H}^{\phi (t)}) U_{N,t} - U_{N,t} W_N^* d\Gamma(H_{H}^{\phi (t)}) W_N\,.
\end{split}
\end{equation}
In the last step, we used that $W_NW_N^*$ acts as the identity in the image of $W_N$, and $d\Gamma(H_{H})$ commutes with both, the number of particles and $W_N$. Multiplying \eqref{eq:U_ndt} by $U_{N,t}^*$ we get the first term of \eqref{eq: L(t)}. For the second term of \eqref{eq: L(t)}, we use $W_N^* d\Gamma(H_N) W_N = H_{N}$, and we thus arrive at 
\begin{equation}\label{eq:L(t)HF}
    \mathcal{L}(t) = d\Gamma(H_{H}) Q_{N,t} + U_{N,t}W_N^*\left( d\Gamma(H_N) -  d\Gamma(H_{H}^{\phi (t)}) \right) W_N  U_{N,t}^*.
\end{equation}
Since the first term commutes with the particle-number operator, it does not contribute to the derivative of $g_N (t;\beta)$ (see \eqref{eq:parital_tg_N}). We compute the second term, based on the representations \eqref{eq:dGamma(H_N)} of $d\Gamma (H_N)$ and $d\Gamma ( H_H^{\phi (t)})$ and the properties of the excitation map \eqref{eq:U_NConjugation}. Using that 
\begin{align}
    &U_{N,t} W_{N}^* a_m^* a_n^* a_q a_p W_{N} U_{N,t}^*\nonumber\\
    &=U_{N,t} W_{N}^* a_m^* W_{N-1}U_{N-1,t}^*U_{N-1,t}W_{N-1}^* a_n^* W_{N-2}U_{N-2,t}^*U_{N-2,t}W_{N-2}^* a_q W_{N-1}U_{N-1,t}^*U_{N-1,t}W_{N-1}^* a_p W_{N}U_{N,t}^*
\end{align}
and the notation $\sum_m {} = \sum_{m\in\mathbb{N}}$ and $\sum_m {}^{'} = \sum_{m\in\mathbb{N}_+}$, we arrive after a straightforward computation at

\begin{equation}
\begin{split}
    &U_{N,t}W_N^* \left(d\Gamma(H_N)-d\Gamma(H_{H}^{\phi (t)}) \right) W_N U_{N,t}^* = \\
    &\frac{1}{2(N-1)} Q_N\Bigg(w_{0000} (N-\mathcal{N})(N-\mathcal{N}-1)+ \sum_{mnpq}{}^{'} w_{mnpq} a_m^* a_n^* a_q a_p \\
    &+ \sum_{npq}{}^{'} w_{0npq} \sqrt{N-\mathcal{N}} a_n^* a_q a_p + \sum_{mpq}{}^{'} w_{m0pq} a_m^* \sqrt{N-\mathcal{N}-1} a_q a_p \\
    &+ \sum_{mnq}{}^{'} w_{mn0q} a_m^* a_n^* a_q \sqrt{N-\mathcal{N}}  + \sum_{mnp0}{}^{'} w_{mnp0} a_m^* a_n^*  \sqrt{N-\mathcal{N}-1} a_p\\
    &+ \sum_{pq}{}^{'} w_{00pq}  \sqrt{N-\mathcal{N}}\sqrt{N-\mathcal{N}-1}a_q a_p + \sum_{nq}{}^{'} w_{0n0q} \sqrt{N-\mathcal{N}} a_n^* a_q \sqrt{N-\mathcal{N}}  \\
    & + \sum_{np}{}^{'} w_{0np0} \sqrt{N-\mathcal{N}} a_n^* \sqrt{N-\mathcal{N}-1} a_p + \sum_{mq}{}^{'} w_{m00q} a_m^* \sqrt{N-\mathcal{N}-1} a_q \sqrt{N-\mathcal{N}}\\
    & + \sum_{mp}{}^{'} w_{m0p0} a_m^* (N-\mathcal{N}-1) a_p + \sum_{mn}{}^{'} w_{mn00} a_m^* a_n^* \sqrt{N-\mathcal{N}-1}\sqrt{N-\mathcal{N}} \\
    & + \sum_{m}{}^{'} w_{m000} a_m^* (N-\mathcal{N}-1) \sqrt{N-\mathcal{N}}+ \sum_{n}{}^{'} w_{0n00} \sqrt{N-\mathcal{N}} a_n^* \sqrt{N-\mathcal{N}-1}\sqrt{N-\mathcal{N}} \\
    & + \sum_{p}{}^{'} w_{00p0}\sqrt{N-\mathcal{N}}(N-\mathcal{N}-1) a_p+ \sum_{q}{}^{'} w_{000q} \sqrt{N-\mathcal{N}} \sqrt{N-\mathcal{N}-1} a_q \sqrt{N-\mathcal{N}}\Bigg)Q_N\\
    & - Q_N\Bigg(\sum_{mp}{}^{'} w_{m0p0} a_m^* a_p + w_{0000}(N-\mathcal{N}) + \sum_{m}{}^{'} w_{m000} a_m^* \sqrt{N-\mathcal{N}}  + \sum_{p}{}^{'} w_{00p0} \sqrt{N-\mathcal{N}} a_p\Bigg)Q_N.
\end{split}
\end{equation}
We note that we set $e_0 = \phi (t)$, and thus the coefficients $w_{k_1,k_2,k_3,k_4}$ for $k_i \in \mathbb{N}$ depend on time. We can write the r.h.s. as 
\begin{equation}\label{eq:A_delta-decomposition}
    U_{N,t}W_N^* \left(d\Gamma(H_N)-d\Gamma(H_{H}) \right) W_N U_{N,t}^*  = Q_{N,t} \sum_{\delta=-2}^{2} A_\delta \ Q_{N,t},
\end{equation}
where $A_{\delta}$ contains operators that destroy $\delta$ particles and $A_{-\delta} = A_{\delta}^*$ contains operators that create $\delta$ particles. The terms further simplify by the symmetries under particle permutation of $w$. In fact, since for the derivative of $g_{N} (t,\beta)$ the operator $A_0$, commuting with the number of particle operator, does not contribute, the only relevant terms are 
\begin{align}
 A_1^* =& \frac{1}{(N-1)}\sum_{mnq}{}^{'} w_{mn0q} a_m^*a_n^*a_q \sqrt{N-\mathcal{N}} -\frac{1}{(N-1)} \sum_m {}^{'} w_{m000} a_m^* \mathcal{N}\sqrt{N-\mathcal{N}}\,, \notag \\
  A_2^* =& \frac{1}{2(N-1)} \sum_{mn}{}^{'} w_{mn00} a_m^* a_n^* \sqrt{N-\mathcal{N}-1}\frac{\sqrt{N-\mathcal{N}}}{N-1}\,,
\end{align}
and their adjoins operators.

In this step the Hartree Hamiltonian plays a crucial role: the second term on $A^*_1$ has a contribution of order $N^{3/2}$  coming from the many-body Hamiltonian which cancels when we subtract the contribution from the Hartree  Hamiltonian and the final results has the correct scaling $\sqrt{N}$.  
Thus, we arrive from \eqref{eq:parital_tg_N} at 
\begin{equation}
  \partial_t g_N (t,\beta) = -i \left\langle  [\exp(\beta\mathcal{N}),\mathcal{L}(t)]\right\rangle = \sum_{\delta=1}^2 2 \;\text{Im}\left\langle  [\exp(\beta\mathcal{N}),A_\delta^*]\right\rangle .
\end{equation}
As an easy consequence of the commutation relations, we have $\mathcal{N} A_\delta^* = A_\delta^*(\mathcal{N} +\delta)$ and therefore 
\begin{equation}
    [\exp{(\beta \mathcal{N})},A_\delta^*]  = 2\sinh\left(\frac{\beta}{2}\right) \exp\left(\frac{\beta\mathcal{N}}{2}\right)A_\delta^*\  \exp{\left(\frac{\beta \mathcal{N}}{2}\right)}\; 
\end{equation}
that enables us to write  \eqref{eq:parital_tg_N} explicitly in terms of creation and annihilation operators as 
\begin{equation}\label{eq:gExplicit}
\begin{split}
    \partial_t g(t,\beta) &= 4\sinh\left(\frac{\beta}{2}\right) \text{Im}\left\langle  \sum_{mnq}{}^{'} \frac{w_{mn0q}}{N-1}\exp{\left(\frac{\beta \mathcal{N}}{2}\right)} a_m^*a_n^*a_q \sqrt{N-\mathcal{N}} \exp{\left(\frac{\beta \mathcal{N}}{2}\right)}\right\rangle \\
    &-4\sinh\left(\frac{\beta}{2}\right) \text{Im}\left\langle  \sum_m {}^{'} \frac{w_{m000}}{N-1} \exp{\left(\frac{\beta \mathcal{N}}{2}\right)} a_m^* \mathcal{N} \sqrt{N-\mathcal{N}}\exp{\left(\frac{\beta \mathcal{N}}{2}\right)}\right\rangle \\
    &+2\sinh\left(\beta\right) \text{Im}\left\langle \sum_{mn}{}^{'} \frac{w_{mn00}}{N-1} \exp{\left(\frac{\beta \mathcal{N}}{2}\right)} a_m^* a_n^* \sqrt{N-\mathcal{N}-1}\sqrt{N-\mathcal{N}} \exp{\left(\frac{\beta \mathcal{N}}{2}\right)}\right\rangle.
\end{split}
\end{equation}
Based on this formula, we derive explicit bounds of the form \eqref{eq:general_bound_dtg} in the next step. 

\subsection{Step 2} \label{sec:step2}

While the calculations of the previous section hold for any choice of the interaction potential $w$, the bounds we prove for all terms of \eqref{eq:gExplicit} will depend on the peculiar assumptions on $w$ in \autoref{theo:ExpCondansation} resp. \autoref{theo:ExpCondansation_coulomb}. We prove the two cases in separate subsections below.

\subsubsection{Mean-field models of type (i)}\label{sec:bound_bounded_potentials}

We start with proving that the estimate \eqref{eq:general_bound_dtg} holds true for any bounded $w$ and with the choice $K:= \|w\|$ as in \autoref{theo:ExpCondansation}. 

We estimate all terms of \eqref{eq:gExplicit} separately and start with the first. For this we define the following operators by their matrix elements

\begin{equation}
    W_{mnpq} = (1-\delta_{m0})w_{mnpq} \delta_{p0},\quad \forall\; m,p\in \mathbb{N}, n,q \in \mathbb{N}_+, \label{eq:defW1}
\end{equation}
\begin{equation}
    C_{pqmn} = \frac{1}{N-1}\left\langle\exp{\left(\frac{\beta \mathcal{N}}{2}\right)} U_{N,t} W_N^* a_m^*a_n^*a_q a_p W_NU_{N,t}^* \exp{\left(\frac{\beta \mathcal{N}}{2}\right)}\right\rangle,\quad \forall\; m,p\in \mathbb{N}, n,q \in \mathbb{N}_+. \label{eq:defC1}
\end{equation}
We note that the operator $W$ is bounded, and in particular, its norm is bounded by $\|W\|\leq \|w\|$. The operator $C$ is positive semi-definite.
Indeed, for any vector $(v_{mn})_{m\in\mathbb{N},n\in\mathbb{N}_+,}$ we have
\begin{equation}
\begin{split}
     &\left\langle v, C v\right\rangle =\left\langle \exp{\left(\frac{\beta \mathcal{N}}{2}\right)} U_{N,t} W_N^* \sum_m \sum_n{}^{'}a_m^*a_n^* \sum_p \sum_q{}^{'}a_q a_p  W_NU_{N,t}^*\exp{\left(\frac{\beta \mathcal{N}}{2}\right)}\right\rangle \\
     &= \left\|\sum_p \sum_q{}^{'}a_q a_p  W_NU_{N,t}^*\exp\left(\frac{\beta \mathcal{N}}{2}\right) \mathcal{U}_N(t,0)\Xi_N \right\|^2>0.
\end{split}
\end{equation}
Since we will use the inequality $|\Tr(WC)|\leq \|W\| \Tr(C)$, first we compute the trace of $C$ using \eqref{eq:Num_+ToNum} and \eqref{eq:NumToN}
\begin{equation}
\begin{split}
    \Tr(C) &= \sum_m\sum_n{}^{'}  \frac{1}{N-1}\left\langle\exp{\left(\frac{\beta \mathcal{N}}{2}\right)} U_{N,t} W_N^* a_m^*a_n^*a_n a_m W_NU_{N,t}^* \exp{\left(\frac{\beta \mathcal{N}}{2}\right)}\right\rangle \\
    & = \frac{1}{N-1}\left\langle U_{N,t} W_N^* \sum_m a_m^*\mathcal{N} a_m W_NU_{N,t}^* \exp{\left(\beta \mathcal{N}\right)}\right\rangle\\
    &=\frac{1}{N-1}\left\langle U_{N,t} W_N^* \left(\sum_m{}^{'} a_m^* a_m(\mathcal{N}_+-1) + a_0^*a_0\mathcal{N}_+\right) W_NU_{N,t}^* \exp{\left(\beta \mathcal{N}\right)}\right\rangle \\
    &=\frac{1}{N-1}\left\langle U_{N,t} W_N^* \ \mathcal{N}_+(\mathcal{N}-1)\ W_NU_{N,t}^* \exp{\left(\beta \mathcal{N}\right)}\right\rangle \\
    &= \frac{1}{N-1}\left\langle \mathcal{N}(N-1) \exp(\beta\mathcal{N})\right\rangle\\
    &= \partial_\beta g_N(t,\beta), 
\end{split}
\end{equation}
where in the last step we identified $\partial_\beta g_N(t,\beta) = \left\langle\mathcal{N}\exp(\beta\mathcal{N})\right\rangle$. Using the operators $W$ and $C$ defined in \eqref{eq:defW1} resp. \eqref{eq:defC1}, the first term in equation \eqref{eq:gExplicit} can be bounded by 
\begin{equation}\label{eq:bound1}
\begin{split}
    &4\sinh\left(\frac{\beta}{2}\right) \Big|\left\langle  \sum_{mnq}{}^{'} \frac{w_{mn0q}}{N-1}\exp{\left(\frac{\beta \mathcal{N}}{2}\right)} a_m^*a_n^*a_q \sqrt{N-\mathcal{N}_+} \exp{\left(\frac{\beta \mathcal{N}}{2}\right)}\right\rangle\Big| \\
    &= 4\sinh\left(\frac{\beta}{2}\right) \Big|\left\langle  \sum_{mp}\sum_{nq}{}^{'} (1-\delta_{m0})\frac{w_{mnpq}}{N-1} \delta_{p0}\exp{\left(\frac{\beta \mathcal{N}}{2}\right)} a_m^*a_n^*a_q \sqrt{N-\mathcal{N}_+} \exp{\left(\frac{\beta \mathcal{N}}{2}\right)}\right\rangle \Big|\\
    &= 4\sinh\left(\frac{\beta}{2}\right) \left|\Tr\left(WC\right)\right|\\
    &\leq 4\sinh\left(\frac{\beta}{2}\right) \|w\| \partial_\beta g_N(t,\beta).
\end{split}
\end{equation}
For the second term of equation \eqref{eq:gExplicit}, we define the following operators by their matrix elements 
\begin{equation}
    \tilde W_{mp} = (1-\delta_{m0})\ w_{m0p0}\, \delta_{p0}, \quad\forall\; m,p \in \mathbb{N},
\end{equation}
\begin{equation}
    \tilde C_{pm} =  \frac{1}{N-1}\left\langle \exp\left(\frac{\beta\mathcal{N}}{2}\right) U_NW_N^* a_m^* \mathcal{N}_+ a_p W_NU_N^* \exp\left(\frac{\beta\mathcal{N}}{2}\right)\right\rangle,\quad \forall\; m,p \in \mathbb{N}.
\end{equation}
We note that $\tilde W$ is bounded with norm $\|\tilde W\|\leq \|w\|$ and the operator $\tilde C$ is positive with trace
\begin{equation}
\begin{split}
    \Tr(\tilde C) &=\frac{1}{N-1}\sum_m\left\langle \exp\left(\frac{\beta\mathcal{N}}{2}\right) U_NW_N^* a_m^* \mathcal{N}_+ a_m W_NU_N^* \exp\left(\frac{\beta\mathcal{N}}{2}\right)\right\rangle \\
    &=  \frac{1}{N-1}\left\langle U_NW_N^* \left( \sum_m {}^{'}a_m^* a_m ( \mathcal{N}_+-1) + a_0^*a_0 \mathcal{N}_+\right)W_NU_N^* \exp\left(\beta\mathcal{N}\right)\right\rangle \\
    &=  \frac{1}{N-1}\left\langle U_NW_N^* (\mathcal{N}-1) \mathcal{N}_+ W_NU_N^* \exp\left(\beta\mathcal{N}\right)\right\rangle \\
    &=  \frac{1}{N-1}\left\langle (N-1)\mathcal{N} \exp\left(\beta\mathcal{N}\right)\right\rangle \\
    &= \partial_\beta g_N(t,\beta).
\end{split}
\end{equation}
With these definitions, we estimate the second term of \eqref{eq:gExplicit} by 
\begin{equation}\label{eq:bound2}
\begin{split}
    &4\sinh\left(\frac{\beta}{2}\right) \Big|\left\langle  \sum_m {}^{'} \frac{w_{m000}}{N-1} \exp{\left(\frac{\beta \mathcal{N}}{2}\right)} a_m^* \mathcal{N} \sqrt{N-\mathcal{N}}\exp{\left(\frac{\beta \mathcal{N}}{2}\right)}\right\rangle\Big| \\
    & = 4\sinh\left(\frac{\beta}{2}\right) \Big|\left\langle  \sum_{mp} (1-\delta_{m0})\frac{w_{m0p0}}{N-1} \delta_{p0} \exp{\left(\frac{\beta \mathcal{N}}{2}\right)}U_N W_N^* a_m^* \mathcal{N}_+ a_p W_N U_N^*\exp{\left(\frac{\beta \mathcal{N}}{2}\right)}\right\rangle\Big| \\
    &= 4\sinh\left(\frac{\beta}{2}\right) |\Tr(\tilde W \tilde C)|\\
    &\leq 4\sinh\left(\frac{\beta}{2}\right) \|w\| \partial_\beta g_N(t,\beta).
\end{split}
\end{equation}

For the third term of equation \eqref{eq:gExplicit}, we will use a different bound: we define the following vectors with entries
\begin{equation}
    v_{mn} = w_{mn00}\quad \forall\; m,n \in \mathbb{N}_+,
\end{equation}
\begin{equation}
    u_{mn} = \frac{1}{N-1}\left\langle \exp{\left(\frac{\beta \mathcal{N}}{2}\right)} a_m^* a_n^* \sqrt{N-\mathcal{N}-1}\sqrt{N-\mathcal{N}} \exp{\left(\frac{\beta \mathcal{N}}{2}\right)}\right\rangle \quad \forall\; m,n \in \mathbb{N}_+.
\end{equation}
Since $w$ is bounded, we have 
    \begin{equation}
        \sqrt{\sum_{mn}|w_{mnpq}|^2} \leq \|w\| \quad\forall\; p,q\in\mathbb{N}\,,
    \end{equation}
and furthermore
\begin{equation}
    \|v\|^2 = \sum_{mn}{}^{'} |w_{mn00}|^2 \leq \|w\|^2.
\end{equation}
To compute the norm of the vector $u$, we artificially add $1 = (\mathcal{N}+1)^{-1/2}(\mathcal{N}+1)^{1/2}$ and estimate with the Cauchy--Schwarz inequality 
\begin{equation}
\begin{split}
    \|u\|^2 &= \frac{1}{(N-1)^2} \sum_{mn}{}^{'}\left|\left\langle \exp{\left(\frac{\beta \mathcal{N}}{2}\right)} a_m^* a_n^* (\mathcal{N}+1)^{-1/2}(\mathcal{N}+1)^{1/2}\sqrt{N-\mathcal{N}-1}\sqrt{N-\mathcal{N}} \exp{\left(\frac{\beta \mathcal{N}}{2}\right)}\right\rangle\right|^2 \\
    &\leq\frac{1}{(N-1)^2}  \sum_{mn}{}^{'}\left\langle \exp{\left(\frac{\beta \mathcal{N}}{2}\right)} a_m^* a_n^*(\mathcal{N}+1)^{-1} a_na_m \exp{\left(\frac{\beta \mathcal{N}}{2}\right)} \right\rangle\\
    &\quad \quad \times\left\langle(\mathcal{N}+1)(N-\mathcal{N}-1)(N-\mathcal{N}) \exp{\left(\beta \mathcal{N}\right)}\right\rangle \\
    &\leq \left\langle \sum_{mn}{}^{'}a_m^* a_n^* a_na_m(\mathcal{N}-1)^{-1} \exp{\left(\beta \mathcal{N}\right)} \right\rangle\left\langle(\mathcal{N}+1)\frac{(N-\mathcal{N}-1)(N-\mathcal{N})}{(N-1)^2} \exp{\left(\beta \mathcal{N}\right)}\right\rangle\\
    &\leq\left\langle\mathcal{N} \exp{\left(\beta \mathcal{N}\right)}\right\rangle\left\langle(\mathcal{N}+1) \exp{\left(\beta \mathcal{N}\right)}\right\rangle\\
    &\leq\left\langle(\mathcal{N}+1)\exp(\beta\mathcal{N})\right\rangle^2.
\end{split} 
\end{equation}
With these notations and bounds, we get for the third term of  \eqref{eq:gExplicit}  
\begin{equation}\label{eq:bound3}
\begin{split}
    &2\sinh\left(\beta\right) \left|\left\langle \sum_{mn}{}^{'} \frac{w_{mn00}}{N-1} \exp{\left(\frac{\beta \mathcal{N}}{2}\right)} a_m^* a_n^* \sqrt{N-\mathcal{N}-1}\sqrt{N-\mathcal{N}} \exp{\left(\frac{\beta \mathcal{N}}{2}\right)}\right\rangle\right| \\
    & = 2\sinh(\beta) |\left\langle v,u\right\rangle | \\
    & \leq 2\sinh(\beta) \|v\|\|u\| \\
    & \leq 2\sinh(\beta) \|w\| (g_N(t,\beta) + \partial_\beta g_N(t,\beta) ).
\end{split}
\end{equation}
Summing up the three bounds \eqref{eq:bound1}, \eqref{eq:bound2} and \eqref{eq:bound3}, we get as a final bound for  \eqref{eq:gExplicit}
\begin{equation}
    \partial_t g_N(t,\beta) \leq  \left( 8\|w\|\sinh\left(\frac{\beta}{2}\right) + 2\|w\| \sinh(\beta)\right) \partial_\beta g_N(t,\beta) + 2\sinh(\beta) \|w\| g_N(t,\beta)\,,
\end{equation}
that proves the claim. 

\subsubsection{Mean-field models of type (ii)}\label{sec:bounds_coulomb_potential}

Now we will prove the claim \eqref{eq:general_bound_dtg} under the  assumptions of \autoref{theo:ExpCondansation_coulomb}. We start by rewriting \eqref{eq:gExplicit} using that in this case $w$ is a multiplication operator. Since the interaction is given by a multiplication operator of the form $w(x;y) = V(x-y)$, we have 
    \begin{equation}\label{eq:w_matrix_elements_multiplication}
    w_{mnpq} = \int dxdy \bar e_m(x)\bar e_n(y)V(x-y) e_p(x) e_q(y).
    \end{equation} 
We observe that, formally, since the expectation value in  \eqref{eq:gExplicit} is evaluated in wavefunctions on the orthogonal Fock space $\mathcal{F}_{\perp \phi (t)}^{\leq N}$, we can replace $\sum'$ by $\sum$ in all three terms of \eqref{eq:gExplicit}. Therefore, using the identity $\sum_n \bar e_n(x)e_n(y) = \delta(x-y)$, the linearity $\sum_n \alpha_n a^*_n = a^*(\sum_n \alpha_n e_n)$, and 
\begin{equation}\label{def:a^*_x}
    \sum_n \bar e_n(x)a^*_n = a^*(\sum_n e_n(x) e_n) = a^*(\delta(x-\cdot )) =: a^*_x\ .
\end{equation}
we can write \eqref{eq:gExplicit} as 
\begin{equation}\label{eq:gExplicit_multiplication}
\begin{split}
    \partial_t g(t,\beta) &= 4\sinh\left(\frac{\beta}{2}\right) \text{Im} \left\langle  \int dxdy \frac{V(x-y)}{N-1}\phi(t,x)\exp{\left(\frac{\beta \mathcal{N}}{2}\right)} a_x^*a_y^*a_y \sqrt{N-\mathcal{N}} \exp{\left(\frac{\beta \mathcal{N}}{2}\right)}\right\rangle \\
    &-4\sinh\left(\frac{\beta}{2}\right) \text{Im} \left\langle  \int dxdy \frac{V(x-y)}{N-1} \bar\phi(t,y)\phi(t,x)\phi(t,y) \exp{\left(\frac{\beta \mathcal{N}}{2}\right)} a_x^* \mathcal{N} \sqrt{N-\mathcal{N}}\exp{\left(\frac{\beta \mathcal{N}}{2}\right)}\right\rangle \\
    &+2\sinh\left(\beta\right) \text{Im}\left\langle \int dxdy \frac{V(x-y)}{N-1} \phi(t,x)\phi(t,y) \exp{\left(\frac{\beta \mathcal{N}}{2}\right)} a_x^* a_y^* \sqrt{N-\mathcal{N}-1}\sqrt{N-\mathcal{N}} \exp{\left(\frac{\beta \mathcal{N}}{2}\right)} \right\rangle.
\end{split}
\end{equation} 
To bound each of these terms we use that the creation and annihilation operators are bounded in terms of the number of particles operator by 
\begin{equation}\label{eq:bounda}
    \|a^*(f) \Psi\| \leq \|f\|_{L^2}\|\sqrt{\mathcal{N}+1}\Psi\|, \quad \|a(f)\Psi\| \leq \|f\|_{L^2} \|\sqrt{\mathcal{N}}\Psi\| \quad \forall\;\Psi\in\mathcal{F}\,,
\end{equation}
valid for all $\Psi \in \mathcal{F}$. On the orthogonal Fock space $\mathcal{F}_{\perp \phi (t)}^{\leq N}$ we furthermore have 
\begin{equation}\label{eq:boundN}
    \|\mathcal{N} \Psi\| \leq N\|\Psi\| \quad \forall\;\Psi \in \F^{\leq N}.
\end{equation}
Moreover, we use that  $\|\phi(t) \|_{L^2}=1\ \forall\; t$ and the constant $\mathbb{V}$ defined in \eqref{def:K} can be written as
\begin{equation}
\mathbb{V} = \sup_x \|V(x-\cdot)\phi(t)\|_{L^{2}}\,.  
\end{equation}
Now we bound each term in \eqref{eq:gExplicit_multiplication} separately using that $\int f(x) a^*_x dx = a^*(f)$, the Cauchy-Schwarz inequality and the inequalities \eqref{eq:bounda}, \eqref{eq:boundN}. 

We start with the first term that we estimate with 
\begin{equation}\label{eq:boundC1}
\begin{split}
    &\left|4\sinh\left(\frac{\beta}{2}\right) \text{Im} \left\langle  \int dxdy \frac{V(x-y)}{N-1}\phi(t,x)\exp{\left(\frac{\beta \mathcal{N}}{2}\right)} a_x^*a_y^*a_y \sqrt{N-\mathcal{N}} \exp{\left(\frac{\beta \mathcal{N}}{2}\right)} \right\rangle \right| \\
    & \leq \frac{4}{N-1}\sinh\left(\frac{\beta}{2}\right)   \int dy  \left| \left\langle\exp{\left(\frac{\beta \mathcal{N}}{2}\right)} a_y^*a^*(V(y-\cdot)\phi (t))\sqrt{N-\mathcal{N}-1}\ a_y \exp{\left(\frac{\beta \mathcal{N}}{2}\right)} \right\rangle \right| \\
    & \leq \frac{4}{N-1}\sinh\left(\frac{\beta}{2}\right)   \int dy  \|a(V(y-\cdot)\phi(t)) a_y \exp{\left(\frac{\beta \mathcal{N}}{2}\right)} \mathcal{U}_N(t,0) \Xi_N \| \\
    &\hspace{10em} \times \|\sqrt{N-\mathcal{N}-1}\ a_y \exp{\left(\frac{\beta \mathcal{N}}{2}\right)} \mathcal{U}_N(t,0) \Xi_N \| \\
    & \leq \frac{4}{N-1}\sinh\left(\frac{\beta}{2}\right)   \int dy  \mathbb{V}(N-1) \| a_y \exp{\left(\frac{\beta \mathcal{N}}{2}\right)} \mathcal{U}_N(t,0) \Xi_N \|^2 \\
    & \leq 4\mathbb{V}\sinh\left(\frac{\beta}{2}\right) \left\langle\mathcal{N} \exp{\left(\beta \mathcal{N}\right)} \right\rangle  =4\mathbb{V}\sinh\left(\frac{\beta}{2}\right) \partial_\beta g_N(t,\beta)\,.
\end{split}
\end{equation}
With similar ideas, we bound the second term by 
\begin{equation}\label{eq:boundC2}
\begin{split}
    &\left|4\sinh\left(\frac{\beta}{2}\right) \text{Im}\left\langle  \int dxdy \frac{V(x-y)}{N-1} \bar\phi(t,y)\phi(t,x)\phi(t,y) \exp{\left(\frac{\beta \mathcal{N}}{2}\right)} a_x^* \mathcal{N} \sqrt{N-\mathcal{N}}\exp{\left(\frac{\beta \mathcal{N}}{2}\right)}\right\rangle \right|\\
    &\leq \frac{4}{N-1}\sinh\left(\frac{\beta}{2}\right) \int dy |\phi(t,y)|^2 \left| \left\langle    \exp{\left(\frac{\beta \mathcal{N}}{2}\right)} a^*(V(y-\cdot)\phi (t)) \mathcal{N} \sqrt{N-\mathcal{N}}\exp{\left(\frac{\beta \mathcal{N}}{2}\right)}\right\rangle \right|\\
    &\leq \frac{4}{N-1}\sinh\left(\frac{\beta}{2}\right) \int dy |\phi(t,y)|^2 \|\sqrt{\mathcal{N}} a(V(y-\cdot)\phi (t)) \exp{\left(\frac{\beta \mathcal{N}}{2}\right)} \mathcal{U}_N (t,0) \Xi_N \| \\
    &\hspace{13em} \times\|\sqrt{\mathcal{N}} \sqrt{N-\mathcal{N}} \exp{\left(\frac{\beta \mathcal{N}}{2}\right)} \mathcal{U}_N (t,0) \Xi_N \| \\
    &\leq \frac{4}{N-1}\sinh\left(\frac{\beta}{2}\right)  \mathbb{V} N \|\sqrt{\mathcal{N}}\exp{\left(\frac{\beta \mathcal{N}}{2}\right)} \mathcal{U}_N (t,0) \Xi_N \|^2 \\
    & \leq 4 \mathbb{V}\frac{N}{N-1} \sinh\left(\frac{\beta}{2}\right)  \left\langle \mathcal{N} \exp{\left(\beta \mathcal{N}\right)}\right\rangle )\\
    & \leq 4 \mathbb{V}(1+\epsilon) \sinh\left(\frac{\beta}{2}\right) \partial_\beta g_N(t,\beta)\,.
\end{split}
\end{equation}
We remark that the last step holds true for any $\epsilon> 1/(N-1)$. Lastly, we bound  the third term by 
\begin{equation}\label{eq:boundC3}
\begin{split}
    &\left| 2\sinh\left(\beta\right) \text{Im} \left\langle \int dxdy \frac{V(x-y)}{N-1} \phi(t,x)\phi(t,y) \exp{\left(\frac{\beta \mathcal{N}}{2}\right)} a_x^* a_y^* \sqrt{N-\mathcal{N}-1}\sqrt{N-\mathcal{N}} \exp{\left(\frac{\beta \mathcal{N}}{2}\right)} \right\rangle \right| \\
    & \leq \frac{2}{N-1}\sinh\left(\beta\right)  \int dx \left| \phi(t,x) \left\langle \exp{\left(\frac{\beta \mathcal{N}}{2}\right)} a_x^* \sqrt{N-\mathcal{N}} a^*(V(x-\cdot)\phi (t)) \sqrt{N-\mathcal{N}} \exp{\left(\frac{\beta \mathcal{N}}{2}\right)} \right\rangle \right| \\
    & \leq \frac{2}{N-1}\sinh\left(\beta\right)  \int dx  |\phi(t,x)| \| \sqrt{N-\mathcal{N}} a_x \exp{\left(\frac{\beta \mathcal{N}}{2}\right)} \mathcal{U}_N(t,0) \Xi_N \| \\
    &\hspace{10em} \times \|  a^*(V(x-\cdot)\phi (t)) \sqrt{N-\mathcal{N}} \exp{\left(\frac{\beta \mathcal{N}}{2}\right)} \mathcal{U}_N(t,0) \Xi_N \| \\
    & \leq \frac{2}{N-1}\sinh\left(\beta\right) \| \sqrt{N-\mathcal{N}} a(\phi (t)) \exp{\left(\frac{\beta \mathcal{N}}{2}\right)} \mathcal{U}_N(t,0) \Xi_N \|\ \\
    &\hspace{10em} \times \mathbb{V} \|  \sqrt{\mathcal{N}+1} \sqrt{N-\mathcal{N}} \exp{\left(\frac{\beta \mathcal{N}}{2}\right)} \mathcal{U}_N(t,0) \Xi_N \|\\
    & \leq 2 \mathbb{V} \frac{N}{N-1}\sinh\left(\beta\right) \|  \sqrt{\mathcal{N}+1} \exp{\left(\frac{\beta \mathcal{N}}{2}\right)} \mathcal{U}_N(t,0) \Xi_N \|^2 \\
    & \leq 2 \mathbb{V}(1+\epsilon) \sinh(\beta) \left(\partial_\beta g_N(t,\beta) + g_N(t,\beta) \right)\,.
\end{split}
\end{equation}

To simplify the notation we multiply the bound on \eqref{eq:boundC1} by $1+\epsilon$. Furthermore, since $N\geq2$, we fix $\epsilon=1$. 
Summing up the three bounds \eqref{eq:boundC1}, \eqref{eq:boundC2} and \eqref{eq:boundC3}, we arrive at 
\begin{equation}
    \partial_t g_N(t,\beta) \leq  \left( 16\mathbb{V}\sinh\frac{\beta}{2} + 4\mathbb{V} \sinh\beta\right) \partial_\beta g_N(t,\beta) + 4\sinh\beta \mathbb{V} g_N(t,\beta)\,,
\end{equation}
that proves the desired claim. 

\subsection{Step 3} \label{sec:step3}

In the last step, we prove the final estimate of \autoref{theo:ExpCondansation_coulomb} resp. \autoref{theo:ExpCondansation} from \eqref{eq:general_bound_dtg} based on a Gronwall type argument. Since this argument does not depend on the choice of $K$ in the estimate \eqref{eq:general_bound_dtg}, this part of the proof is again the same for both Theorems.

To simplify the calculations, we first bound
\begin{equation}
    8K\sinh\frac{\beta}{2} + 2K \sinh\beta \leq 6K \sinh\beta.
\end{equation}
and write  \eqref{eq:general_bound_dtg} as 
\begin{equation}
    \frac{1}{2K\sinh\beta}\partial_t g_N(t,\beta) + 3 \partial_\beta g_N(t,\beta) \leq g_N(t,\beta).
\end{equation}
In particular note that all coefficients in front of (derivatives of) $g_N (t,\beta)$ are independent of $N$.  However, before we can apply Gronwall's lemma, we need to perform a suitable change of variables $x=X(t,\beta)$ and $y=Y(t,\beta)$ such that the inequality takes the form 
\begin{equation}
\label{eq:GW}
    \partial_y g_N \leq g_N\,.
\end{equation}
In fact, from \eqref{eq:GW}, we then arrive for fixed $x$ from Gronwall's inequality at 
\begin{equation}\label{eq:Gronwall}
    g_N(x,y) \leq g_N(x,y_0) e^{y-y_0} 
\end{equation}
for arbitrary $y_0 \in \mathbb{R}$. We define $y_0$ through the inverse change of variables $t = T(x,y)$ and $\beta = B(x,y)$ such that it realizes the initial conditions $g_N(t=0,\beta) \leq C_\beta$, i.e. we define $y_0$ through 
\begin{equation}\label{def:y_0}
    T(X(t,\beta),y_0) = 0,
\end{equation}
such that $g_N(x,y_0)\leq C_\beta$. Thus, the change of variables $(X,Y)$, solves 
\begin{equation}
    \frac{1}{2K\sinh\beta} \partial_tX-3\partial_\beta X = 0,  \quad \text{and} \quad  \frac{1}{2K\sinh\beta} \partial_tY-3\partial_\beta Y = 1 \; . 
\end{equation}
The functions 
\begin{equation}\label{eq:X_explicit}
    X(t,\beta) = 6Kt  - 2 \atgh(e^{-\beta}) , \quad \text{and} \quad   Y(t,\beta) =  6Kt - 2 \atgh(e^{-\beta}) -\frac{\beta}{3}
\end{equation}
solve the equations and any other solution is a reparameterization of this variables and will therefore not change the result. The inverse change of variables is given by
\begin{equation}\label{eq:T_explicit}
    T(x,y) = \frac{1}{6K}\left(x + 2 \atgh\left(e^{-3\left(x-y\right)}\right) \right), \quad \text{and} \quad  B(x,y)=3(x-y).
\end{equation}
Since $\beta\geq0$ by assumption, we find $x-y \geq 0$. From \eqref{def:y_0} we finally get 
\begin{equation}\label{eq:y_0_explicit}
    y_0(t,\beta) = 6Kt - 2 \atgh(e^{-\beta}) + \frac{1}{3} \ln\left(\tanh\left( \atgh(e^{-\beta})- 3Kt\right)\right)
\end{equation}
for all 
\begin{equation}
    0\le\beta<\beta_c(t) := -\ln\tanh(3Kt)\,,
\end{equation}
that ensures the argument of the logarithm is positive. Thus the initial condition $g_N(x,y_0)\leq C_\beta$ is realized only for $\beta < \beta_c(t)$. With the choices \eqref{eq:X_explicit} and \eqref{eq:y_0_explicit} in \eqref{eq:Gronwall} we finally arrive at 
\begin{equation}
    g_N(t,\beta) \leq C_\beta \left( \frac{1-\tanh(3Kt)e^{-\beta}}{1-\tanh(3Kt)e^{\beta}}\right)^{\frac{1}{3}}.
\end{equation}

\section{Conclusions}

In this work we have analyzed the persistence of Bose–Einstein condensation for bosonic many-body systems evolving under mean-field dynamics, with a particular focus on the quantitative control of the number of excitations above the condensate. Starting from initial data exhibiting an exponential concentration of the number of excitations, we proved that this property is preserved for all finite times: the probability of finding $n$ particles outside the condensate decays exponentially in $n$, uniformly in the total particle number $N$.

Our results apply to two broad and physically relevant classes of models: systems with arbitrary bounded two-body interactions, encompassing a wide range of finite-dimensional and effective mean-field models, and continuum systems with unbounded interaction potentials satisfying the operator inequality $V^2 \le D(1-\Delta)$, including Coulomb-type interactions. In both settings, the exponential control of the number of excitations significantly sharpens previously available bounds, which were limited to polynomial decay derived from uniform estimates on finitely many moments of the excitation number operator.

From a physical perspective, the exponential concentration of the number of excitations provides a refined description of the many-body state beyond the leading-order Hartree dynamics. This type of control is particularly relevant in situations where rare but large fluctuations may play a role, for instance in the study of correlation functions or in the justification of effective theories for excitations.

On the mathematical side, our results contribute to the growing body of work aimed at a detailed understanding of fluctuations in bosonic mean-field limits. They suggest that, at least for finite times, the excitation structure of the many-body wavefunction is considerably more rigid than what moment bounds alone can capture. It remains an open question to understand if the exponential condensation holds true also in the Gross-Pitaevskii scaling regime.

\section*{Acknowledgements}
We thanks Niels Benedikter for helpful discussions.
GDP has been supported by the HPC Italian National Centre for HPC, Big Data and Quantum Computing -- Proposal code CN00000013 -- CUP J33C22001170001 and by the Italian Extended Partnership PE01 -- FAIR Future Artificial Intelligence Research -- Proposal code PE00000013 -- CUP J33C22002830006 under the MUR National Recovery and Resilience Plan funded by the European Union -- NextGenerationEU.
Funded by the European Union -- NextGenerationEU under the National Recovery and Resilience Plan (PNRR) -- Mission 4 Education and research -- Component 2 From research to business -- Investment 1.1 Notice Prin 2022 -- DD N. 104 del 2/2/2022, from title ``understanding the LEarning process of QUantum Neural networks (LeQun)'', proposal code 2022WHZ5XH -- CUP J53D23003890006.
GDP and DP are members of the ``Gruppo Nazionale per la Fisica Matematica (GNFM)'' of the ``Istituto Nazionale di Alta Matematica ``Francesco Severi'' (INdAM)''. SR is supported by the European Research
Council via the ERC CoG RAMBAS–Project–Nr. 10104424.

\bibliographystyle{unsrt}
\bibliography{Mibib}

\end{document}